\documentclass[runningheads]{llncs}
\usepackage[T1]{fontenc}
\usepackage{graphicx}
\usepackage{array}
\usepackage{longtable} 

\usepackage{tikz}
\usetikzlibrary{positioning,arrows.meta,fit,backgrounds}

\usepackage{hyperref}
\usepackage[utf8]{inputenc} 
\begin{document}
%

\title{Early Results from Teaching Modelling for Software Comprehension in New-Hire Onboarding}

\titlerunning{Teaching Modelling in New-Hire Onboarding}

\author{Mrityunjay Kumar\inst{1,2}\orcidID{0000-0003-2819-759X} \and
        Venkatesh Choppella\inst{1}\orcidID{0000-0003-1085-3464}}

\institute{
International Institute of Information Technology, Hyderabad 500032, India\\
\email{mrityunjay.k@research.iiit.ac.in}\\
\email{venkatesh.choppella@iiit.ac.in}
\vspace{1em}
\and
Birla Institute of Technology and Science, Pilani, Rajasthan 333031, India\\
\email{mrityunjay.kumar@pilani.bits-pilani.ac.in}
}

%



%
\maketitle              
\begin{abstract}
Working effectively with large, existing software systems requires strong comprehension skills, yet most graduates enter the industry with little preparation for this challenge. We report early results from a pilot intervention integrated into a SaaS company's onboarding program: a five-session course introducing systems thinking and Labelled Transition System (LTS) modelling. Participants articulated their understanding of product behaviour using a structured template and completed matched pre- and post-assessments. Of 35 new hires, 31 provided paired records for analysis. Across the full cohort, gains were small and not statistically significant. However, participants below the median on the pre-test improved by 15 percentage points on average (statistically significant), while those above the median regressed slightly (not statistically significant). Course feedback indicated high engagement and perceived applicability. These results suggest that short, modelling-focused onboarding interventions can accelerate comprehension for less-prepared new hires. At the same time, they point to the need for differentiated pathways for stronger participants, and to the potential for companies to adopt such interventions at scale as a low-cost complement to existing onboarding.

\keywords{System modelling, Software comprehension, Software engineering education, Professional skills, Onboarding, New campus hires, Short Courses}
\end{abstract}

\section{Introduction}
\label{sec:introduction}

Most entry-level software engineers begin with bug fixing and incremental enhancements to existing systems, tasks that demand maintenance and comprehension skills. Yet these skills have received limited attention in both curricula and research. A systematic review of software engineering education by Garousi et al.\ \cite{garousiClosingGapSoftware2020} showed that \emph{Maintenance} was consistently rated among the least important competencies, despite its centrality in practice. Similarly, curricula (SE2014 \cite{ardis_se_2015} and AICTE \cite{jalote_model_2023}) and capstone courses \cite{tenhunen_systematic_2023} give minimal attention to maintenance and evolution.

The mismatch between industry expectations and graduate preparation has been well documented \cite{radermacher_investigating_2014,tuzun_are_2018,kuhrmann_walking_2019}. Employers report that new campus hires often lack practical skills for working with existing codebases and show gaps in teamwork and problem-solving skills \cite{akdur_analysis_2023,watson_attitudes_2017}. Our own surveys of campus hires at two SaaS companies (N=29 and N=15) reinforced this concern: only 20\% expected to reach adequate comprehension by the end of onboarding, while more than half anticipated needing three months or more.

To address this gap, we designed a short, practice-situated course for a SaaS company's product onboarding program. The course introduced systems thinking and modelling, specifically using LTS (Labelled Transition Systems), as strategies for grasping complex software \cite{hmelo-silver_understanding_2006,morecroft_mental_2004}. Learners articulated their understanding of a product using a structured template, which we evaluated for evidence of comprehension.

\vspace{1em}

\textbf{Research Question (RQ):} What is the impact of a short course on systems and modelling on the software comprehension abilities of new campus hires?

\vspace{1em}

We operationalise comprehension as the \emph{articulation of understanding}: the ability to describe system components, behaviours, and interactions in a structured form \cite{goel_towards_1996,choppella_algodynamics_2021}. 

This short paper presents early results from a pilot intervention with new campus hires to improve their comprehension skills. We focused on differential effects of the intervention on below- and above-baseline cohort learners (baseline=pre-test median). 

Onboarding is a critical period when organisations need new campus hires to become productive quickly; 
teaching comprehension skills at this stage can reduce ramp-up time, increase confidence, 
and ease the mentoring load on senior engineers. In addition to contributing to research 
on teaching system comprehension, the study also offers practical insights for organisational 
practice in onboarding.

\section{Related Work}
\label{sec:relatedwork}

Research in science education shows that modelling supports system comprehension. For example, Hmelo-Silver and colleagues, using the Structure-Behavior-Function (SBF) model, observed that novices tend to focus on surface components, whereas experts integrate structural, functional, and behavioural elements \cite{goel_towards_1996,hmelo-silver_understanding_2006,hmelo_silver_expert_2007}; in other words, experts model to understand, novices don't. A similar principle applies to complex software systems; modelling can scaffold novice understanding \cite{hashem_learning_2013}.

Despite modelling being valuable for novice understanding of complex systems, teaching modelling remains peripheral in computing curricula. The ITiCSE 2012 working group found that modelling was often considered a graduate-level topic, lacked a clear definition, and was insufficiently integrated across undergraduate programs. They recommended introducing modelling early and embedding it throughout the curriculum \cite{borstler_teaching_2012}. Research reports on maintenance and evolution courses are limited, and even those that exist \cite{CSSE375,buchtaTeachingEvolutionOpenSource2006,gokhaleTeachingSoftwareEngineering2013a,gallagherTeachingSoftwareMaintenance2019a,hebig_how_2020} rarely emphasise modelling as a comprehension tool.

A detailed review\footnote{\url{https://bit.ly/c25-nirf}} of curricula in leading Indian institutes (top 25 NIRF-ranked) and the AICTE model curriculum further confirmed this gap: software engineering courses typically emphasise lifecycle activities and software creation, with limited coverage of comprehension or modelling. Where modelling appears, it is usually restricted to object-oriented design or formal methods electives.

Our work responds to these gaps by designing a short, practice-oriented course that uses LTS to foster comprehension of complex software. The pilot intervention studied here represents a thin slice of a broader semester-long design.

\section{Experiment Design}
\label{sec:design}

\begin{figure}
    \centering
    \includegraphics[width=0.95\linewidth]{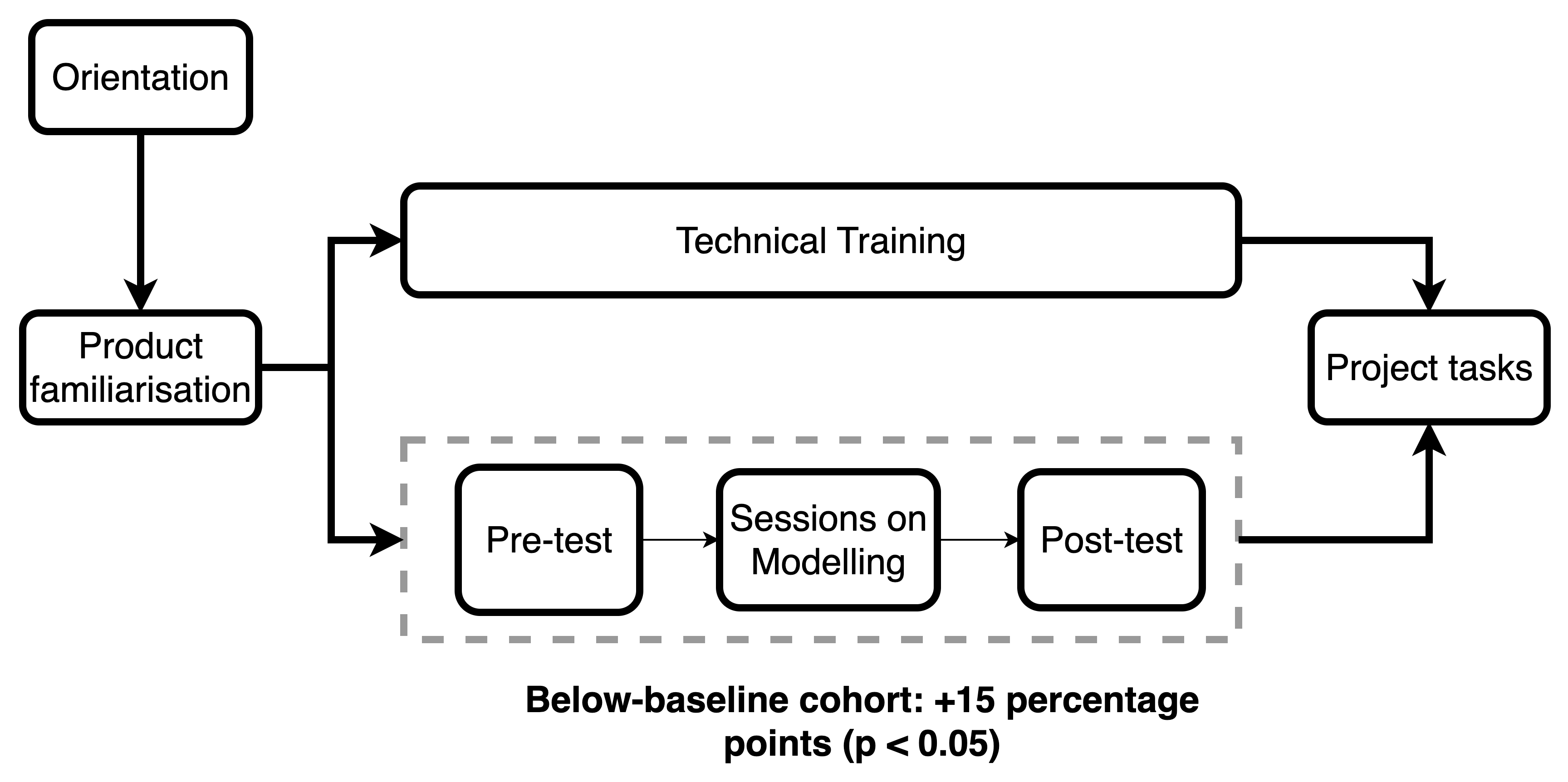}
    \caption{Positioning the intervention within the broader onboarding program. The dashed box shows the study subset, which runs in parallel with technical training; the bold text highlights the key result for the below-baseline cohort (baseline=pre-test median).}
    \label{fig:onboarding-flow}
\end{figure}

\subsection{Context}
This study was conducted with new campus hires (initially hired as interns) at Company~Z, a SaaS product firm with a large India development centre. Each year, the company hires 30-40 final-semester students as interns, who undergo structured technical training and product familiarisation. For this intervention, all 35 interns in the engineering organisation were included; 31 provided complete paired records (pre-test and post-test submissions) and were analysed. The interns were divided into two groups for ease of course delivery. As the intervention was part of the onboarding program, no control group was used and each group was delivered the same content.

\subsection{Intervention and Content}
The short course was designed with two objectives: (i) to help learners view software as a system model, and (ii) to apply LTS modelling to real-world software. It was delivered over five sessions within one month (Fig.~\ref{fig:onboarding-flow}). The opening session (2 hours) introduced the course, explained the articulation template, and administered the pre-test. Three content sessions (2.5 hours each, on alternate days) covered: 
\begin{itemize}
    \item Systems and models, with introductory LTS concepts and examples;
    \item Viewing software as a system and modelling it as an LTS;
    \item Modelling large systems using hierarchical abstractions and a system-of-systems perspective.
\end{itemize}
Examples and exercises were drawn from product modules at Company~Z. The final session (2 hours) concluded the course and administered the post-test. Content was adapted from a semester-long university course but streamlined for integration into the company's onboarding program.

\subsubsection*{Content Details}
The three content sessions combined conceptual input with hands-on activities. Session~1 introduced models and systems, using everyday examples (e.g., a light bulb, Twitter ``heart'' feature) to explain LTS. Session~2 extended this to a product module case study; participants extracted states, actions, and transitions from a simplified description of a membership feature and constructed an LTS. Session~3 addressed large-scale comprehension by decomposing a system (e.g., appointment booking) into subsystems, defining LTS models for each, and articulating their interactions as a system of systems. For instance, learners modelled an expense approval workflow as an LTS with submission, review, and approval states, then extended it to capture rejection and resubmission flows. 

A more detailed supplement with examples and assignments is available online\footnote{\url{https://bit.ly/c25supp}}.

\subsection{Assessment}
We used pre- and post-tests which followed identical format: a product briefing and live demonstration, time for reflection, and 45 minutes for participants to document their understanding using a structured template. Different but complexity-matched SaaS products were used to minimise test-retest effects. 

\noindent \textbf{Scoring rubric.} 
Responses were evaluated against a gold-standard solution created by the instructor. 
The rubric covered four dimensions: (i) correct identification of system components, 
(ii) description of the behaviour of each component, (iii) explanation of interactions 
between components, and (iv) articulation of overall system behaviour in terms of component interactions. 

Each dimension was scored on a 0--5 scale for a maximum of 20 points. Partial credit was awarded for accurate but differently phrased descriptions.

\subsection{Data Collection}
Data comprised (i) Pre-/post-test artefacts scored with the rubric, and (ii) Three Questionnaires (delivered through surveys): Background questionnaire, Topic knowledge questionnaire (post-course knowledge of systems and modelling), and Course feedback questionnaire (perceptions of engagement, skill gains, and applicability). The course feedback used Items adapted from the EDUCATOOL\footnote{\href{https://bit.ly/c25-edutoolkit}{EDUcational Course Assessment TOOLkit (EDUCATOOL)}} toolkit. Participation was voluntary, and consent was obtained.

\section{Results}
\label{sec:results}

\subsection{Background questionnaire}
The responses to background questionnaire ($N=35$) indicated that participants had completed core CS courses and were proficient in common programming languages. However, few had prior internships or substantial project experience, and their exposure to modelling and system-level thinking was limited.

Of the 35 participants, four provided incomplete pre- or post-test submissions. These were excluded, leaving 31 paired records for analysis. All descriptive and inferential results reported below are based on this set ($N=31$).

\subsection{Learning Outcomes}
Paired pre-/post-test data were available for 31 participants. Across the full cohort, mean scores rose only slightly (pre $M=44.5$, post $M=46.9$), a non-significant gain of 2.4 percentage points. When split at the median pre-test score ($Mdn=43.5$), distinct patterns emerged (Table \ref{tab:results}): 

\begin{itemize}
    \item \textbf{Below-baseline cohort ($N=14$)} showed a statistically significant mean gain of 15 percentage points ($p=0.017$).
    \item \textbf{Above-baseline cohort ($N=17$)} exhibited a small, non-significant decline (-8 points).
\end{itemize}
\noindent \textit{Note.} The split was made at the median pre-test score ($Mdn=43.5$). 
With $N=31$, this yielded 14 participants below the median and 17 at or above it; 
ties at the median were grouped with the upper half.

\begin{table}[ht]
\centering
\caption{Pre-/post-test means (M) and gains by subgroup. 
*p<0.05; ns = non-significant. Shapiro-Wilk test confirmed normality assumptions.}
\label{tab:results}
\begin{tabular}{|l|c|c|c|}
\hline
\textbf{Cohort} & \textbf{Pre M (SD)} & \textbf{Post M (SD)} & \textbf{Gain (pp)} \\
\hline
Full cohort ($N=31$) & 44.5 (17.7) & 46.9 (22.7) & +2.4 (ns) \\
Below-baseline cohort ($N=14$) & 29.7 (9.3) & 44.8 (22.4) & +15.1* \\
Above-baseline cohort ($N=17$) & 56.6 (13.1) & 48.7 (23.5) & -8.0 (ns) \\
\hline
\end{tabular}
\end{table}

\begin{figure}[h]
    \centering
    \includegraphics[width=0.9\linewidth]{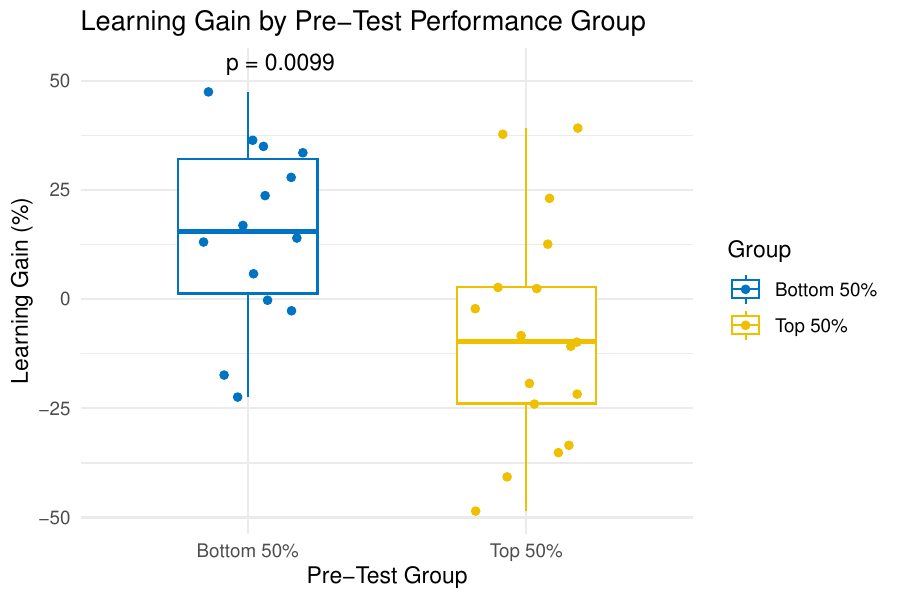}
    \caption{Learning gains by subgroup (below-baseline vs above-baseline, baseline=pre-test median).}
    \label{fig:gainbygroup}
\end{figure}

Fig \ref{fig:gainbygroup} highlights that below-baseline cohort (baseline=pre-test median) improved, while several above-baseline cohort learners regressed slightly. Overall, the intervention appears most beneficial for participants starting with weaker comprehension.

\section{Discussion}
\label{sec:discussions}

The intervention appeared most beneficial for participants who began with lower comprehension. Their significant gains suggest a scaffolding effect \cite{sweller_cognitive_1988,wood_role_1976}, where the articulation template and modelling practice supported learning. This aligns with prior work showing that modelling helps novices reason about system behaviour \cite{hmelo-silver_understanding_2006}. In contrast, above-baseline cohort (baseline=pre-test median) showed no improvement, and some regressed slightly. Possible explanations include the limited scope (five sessions), over-reliance on user interface descriptions rather than system abstractions \cite{vainio_factors_2007}, and potential differences in complexity between the pre- and post-test systems as perceived by the learners.

Course feedback indicated strong acceptance: over 90\% rated the sessions highly for satisfaction, engagement, and applicability. Participants noted a shift in perspective toward systems-level thinking, though some requested more varied examples and activities. This suggests the feasibility and perceived value of incorporating modelling into onboarding, while pointing to design improvements such as differentiated tasks for stronger learners.

\subsection{Practical Implications}
For organisations, these findings highlight that even short, low-cost interventions can help less-prepared new campus hires build the comprehension needed for maintenance and feature work. Embedding such sessions into onboarding could reduce the time taken to achieve productive contribution, and may also improve retention by lowering early frustration. For stronger new campus hires, the results suggest the importance of differentiated challenges, pointing to the value of tiered activities or advanced modules. The intervention required only five sessions, making it a relatively low-cost and logistically feasible complement to existing product onboarding programs.

\subsection{Limitations}

This pilot has several limitations. First, the absence of a control group means that concurrent onboarding activities may also have contributed to observed gains. Second, although the pre- and post-test systems were selected to be comparable, differences in complexity cannot be fully ruled out. Third, comprehension was inferred from written articulations, which may reflect expression skills as much as actual understanding. Fourth, all scoring was performed by a single rater, introducing potential subjectivity despite the structured rubric. Finally, the small sample size ($N=31$) limits generalisability, and subgroup analyses should be interpreted as exploratory.

\section{Conclusion}
\label{sec:conclusion}

Graduating engineers often enter industry with limited preparation for working with large, existing software systems. Curricula tend to emphasise software creation over evolution, leaving few opportunities to practise comprehension at scale. Yet the ability to understand complex systems is a critical prerequisite for almost all engineering work in professional settings. This paper reported on a pilot intervention integrated into the onboarding of new campus hires at Company~Z, a SaaS product company. The short course introduced systems thinking and modelling, supported by pre- and post-tests designed to assess participants' ability to articulate software understanding.

The results show a differential impact. While overall learning gains across the cohort were not statistically significant, the below-baseline cohort (baseline=pre-test median) improved by an average of 15 percentage points, a statistically significant result. This suggests that teaching system modelling can meaningfully support learners with weaker initial comprehension. In contrast, the above-baseline cohort showed no improvement and, in some cases, regressed, indicating that the intervention, in its present form, may not provide sufficient challenge or depth for stronger learners.

These findings demonstrate the promise of incorporating modelling into onboarding, while also underlining that this is work in progress. The study's limitations, including the absence of a control group, single-rater scoring, and the use of different systems for pre- and post-tests, temper the conclusions and highlight the need for further evidence.

Future work will extend the intervention with longer duration and differentiated activities to support advanced learners. We will also incorporate multiple evaluators, explore alternative measures of comprehension, and include control groups to strengthen causal claims. In addition, qualitative methods such as think-aloud protocols and semi-structured interviews will be used to probe how learners actually reason about complex systems. Taken together, these next steps aim to develop a more robust understanding of how modelling can be leveraged to build comprehension skills that are essential for engaging productively with existing software in industry. Beyond research, future work will also focus on designing interventions that companies can adopt at scale. Our findings indicate that teaching system modelling in onboarding is feasible, is valued by learners, and offers organisations a pathway to strengthen new-hire readiness for real-world software work.

\section{Acknowledgements}
\label{sec:acknowledgements}
We used OpenAI ChatGPT-4o (accessed between 20 June 2025 and 27 July 2025) for idea-generation, outlining, and language polishing. All outputs were reviewed and revised by the authors, who take full responsibility for the final content; ChatGPT is not listed as an author.

All data processing and figure generation were performed with \texttt{R}.


\bibliographystyle{splncs04}
\bibliography{prompts}

%




\end{document}